\documentclass[nofootinbib]{revtex4}
\usepackage{amsmath,amssymb,graphicx,epsf}
\usepackage[mathscr]{euscript}
\usepackage{fouriernc}

\begin{document}

\def\babc{\begin{subequations}}
\def\eabc{\end{subequations}}
\def\be{\begin{equation}}
\def\ee{\end{equation}}
\def\ba{\begin{array}}
\def\ea{\end{array}}
\def\nn{\nonumber}
\def\hh{\hspace*{0.25mm}}
\def\h{\hspace*{0.5mm}}
\def\ua{\uparrow}
\def\da{\downarrow}
\def\2{{\textstyle\frac12}}
\def\32{{\textstyle\frac32}}

\title{Boundary Conditions and Formation of Pure Spin Currents in Magnetic Field}

\author{Merab Eliashvili$^{1,2}$ and George Tsitsishvili$^{1,2}$\footnote{giorgi.tsitsishvili@tsu.ge}}
\affiliation{
$^1$Department of Physics, Tbilisi State University, Chavchavadze Ave. 3, Tbilisi 0179, Georgia\\
$^2$Razmadze Mathematical Institute, Tbilisi State University, Tamarashvili Str. 6, Tbilisi 0177, Georgia}

\begin{abstract}
Schr\"odinger equation for an electron confined to a two-dimensional strip is considered in the presence of homogeneous
orthogonal magnetic field. Since the system has edges, the eigenvalue problem is supplied by the boundary conditions (BC)
aimed in preventing the leakage of matter away across the edges. In the case of spinless electrons the Dirichlet and Neumann
BC are considered. The Dirichlet BC result in the existence of charge carrying edge states. For the Neumann BC each separate
edge comprises two counterflow sub-currents which precisely cancel out each other provided the system is populated by
electrons up to certain Fermi level. Cancelation of electric current is a good starting point for developing the spin-effects.
In this scope we reconsider the problem for a spinning electron with Rashba coupling. The Neumann BC are replaced by Robin BC.
Again, the two counterflow electric sub-currents cancel out each other for a separate edge, while the spin current survives thus
modeling what is known as pure spin current -- spin flow without charge flow.
\end{abstract}

\maketitle

\section*{1. Introduction}

The standard notion of electric current implies the directional flow of electrons with no preferred spin orientation.
This results into the charge current with vanishing net spin flow. If electron spins are correlated for certain
reasons, then alongside with the electric current one observes what is known as the spin current \cite{mvsk,mauis,an}.
Considerable amount of studies \cite{islam,gotte,li,nita,linder,futterer,gong,huang,frolov,scheid,shi} are devoted to
the issue of the pure spin current -- the flow of electron spin without flow of electric charge.

Schr\"odinger equation for an electron confined to a two-dimensional strip is considered in the presence of homogeneous
orthogonal magnetic field. It is shown that in the case of spinning electrons with Rashba spin-orbit interaction, the Robin
boundary conditions (BC) imposed on the wave function along the edges produce pure spin currents. For the sake of clarity
we start with spinless electrons in Section 2 and point out the difference between the dispersion relations produced by the
Dirichlet and Neumann BC. In Section 3 we discuss the electric currents carried by edge states and show that for Neumann
BC each of the two edges accommodates two counterflow electric currents which precisely cancel out each other, {\it i.e.}
the electric conductance of a separate edge is zero. In Section 4 we reconsider the problem for spinning electrons with
Rashba spin-orbit interaction. In that case the Neumann BC are replaced by the Robin BC, leading to the same conclusion regarding
the precise cancellation of electric currents at each edge separately. In contrast, the spin current is found to be finite,
meaning the occurrence of pure spin current.

\section*{2. Spinless electron in homogeneous orthogonal magnetic field}

Quantum mechanical Hamiltonian is given by
\be
H=\frac{1}{2m}(i\hbar\partial_n+eA_n)^2,
\ee
where $A_n$ is the vector potential with $B=\partial_xA_y-\partial_yA_x$.

We study the system with the geometry of infinite length $-\infty<y<+\infty$ and finite width $x_L\leqslant x\leqslant x_R$
with $x_L=-\2d$ and $x_R=+\2d$. Correspondingly, solving the eigenvalue problem, the wave function $\psi(x,y)$ has to be
exposed to some boundary conditions (BC) preventing the leakage of a matter across the edges.

The matter flow is described by matter currents
\be
\mathscr J_n=\frac{1}{2im}\big[\psi^\dag(\hbar\partial_n\psi-ieA_n\psi)-(\hbar\partial_n\psi-ieA_n\psi)^\dag\psi\big].
\ee
Then the BC imposed on $\psi(x,y)$ must guarantee vanishing of $x$-component of the current (2) at boundaries
\be
\mathscr J_x(x_L,y)=\mathscr J_x(x_R,y)=0.
\ee

These conditions can be realized in a different ways, and here we comment on the following two options.
One is the Dirichlet BC
\be
\psi(x_L,y)=\psi(x_R,y)=0,
\ee
and the other one is the Neumann BC
\be
\partial_x\psi(x_L,y)=\partial_x\psi(x_R,y)=0.
\ee

Both of these options reproduce (3), but lead to significantly distinct dispersion relations,
hence to distinct physical outcomes. In order to make this statement clear we pass to solving
the eigenvalue problem.

Usage of the Landau gauge $\boldsymbol{A}=(0,Bx)$ secures translational
invariance of the Hamiltonian in $y$-direction. Then the wave function can be written as
\be
\psi(x,y)=e^{+iky}\phi_k(\xi),
\ee
where $k$ is the momentum, and $\xi\equiv\ell^{-1}x+k\ell$ with $\ell$ being the magnetic length
set by ($eB<0$ is assumed)
\be
\frac{1}{\ell^2}=-\frac{eB}{\hbar}.
\ee

By use of (6) the aforementioned boundary conditions are reformulated in terms of $\phi_k(\xi)$ and appear as
\babc
\begin{align}
\textrm{Dirichlet BC :\hspace*{10mm}$\phi_k(\xi_L)=\phi_k(\xi_R)=0$},\\
\nn\\
\textrm{Neumann BC :\hspace*{10mm}$\phi'_k(\xi_L)=\phi'_k(\xi_R)=0$},
\end{align}
\eabc
where
\babc
\begin{align}
\xi_L\equiv-\2\ell^{-1}d+k\ell,\\
\nn\\
\xi_R\equiv+\2\ell^{-1}d+k\ell.
\end{align}
\eabc

The eigenvalue problem for $H$ is reduced to the equation $\mathscr H_k\phi_k(\xi)=\epsilon(k)\phi_k(\xi)$ where
\be
\mathscr H_k=-\2\hh\partial_\xi^2+\2\hh\xi^2.
\ee

Parameterizing eigenvalues as $\epsilon=\nu+\frac12$ the general solution appears as
\be
\phi_k(\xi)=e^{-\frac12\xi^2}\Big[c_1M\big(-\2\nu,\2,\xi^2\big)+c_2\xi M\big(\2-\2\nu,\textstyle{\frac32},\xi^2\big)\Big],
\ee
where $M(a,b,z)$ is the Kummer function, and the constants $c_{1,2}$ to be determined by boundary and normalization conditions.

Consider first the Dirichlet BC (8a). Using (11) these appear as
\babc
\begin{align}
c_1M\big(-\2\nu,\2,\xi_L^2\big)+c_2\xi_LM\big(\2-\2\nu,\textstyle{\frac32},\xi_L^2\big)=0,\\
\nn\\
c_1M\big(-\2\nu,\2,\xi_R^2\big)+c_2\xi_RM\big(\2-\2\nu,\textstyle{\frac32},\xi_R^2\big)=0.
\end{align}
\eabc

This system has nontrivial solution for $c_{1,2}$ only if the corresponding determinant vanishes.
Employing the Kummer transformation $M(a,b,z)=e^zM(b-a,a,-z)$ this condition can be expressed as%
\footnote{Reasoning for Kummer transformation:
increasing $\xi^2$, the value of $M(a,b,\xi^2)$ becomes exponentially large, while $M(a,b,-\xi^2)$
does not, hence more appropriate for numeric calculations.}
\be
f_D(\xi_L)=f_D(\xi_R),
\ee
where
\be
f_D(\xi)\equiv\frac{M\big(\2+\2\nu,\2,-\xi^2\big)}{\xi M\big(1+\2\nu,\textstyle{\frac32},-\xi^2\big)}.
\ee

Eq. (12) determines $\nu$ as a function of $k$, {\it i.e.} the dispersion $\epsilon(k)=\nu(k)+\frac12$.
Solving (13) numerically one obtains the dispersion law shown in the left panel of figure 1.
It should be noted that the dispersion curves produced by Dirichlet BC have been discussed in \cite{halperin}.

Consider now the Neumann BC, which by use of (11) is brought to the form
\be
f_N(\xi_L)=f_N(\xi_R),
\ee
where
\be
f_N(\xi)\equiv
\frac{\xi M\big(\textstyle{\frac12}+\textstyle{\frac12}\nu,\textstyle{\frac12},-\xi^2\big)+2\nu\xi M\big(\textstyle{\frac12}+\textstyle{\frac12}\nu,\textstyle{\frac32},-\xi^2\big)}
{(1-\xi^2)M\big(1+\textstyle{\frac12}\nu,\textstyle{\frac32},-\xi^2\big)+\textstyle{\frac23}(1-\nu)\xi^2M\big(1+\textstyle{\frac12}\nu,\textstyle{\frac52},-\xi^2\big)}.
\ee
The corresponding curve takes the shape shown in the right panel of figure 1.

\begin{center}
\includegraphics{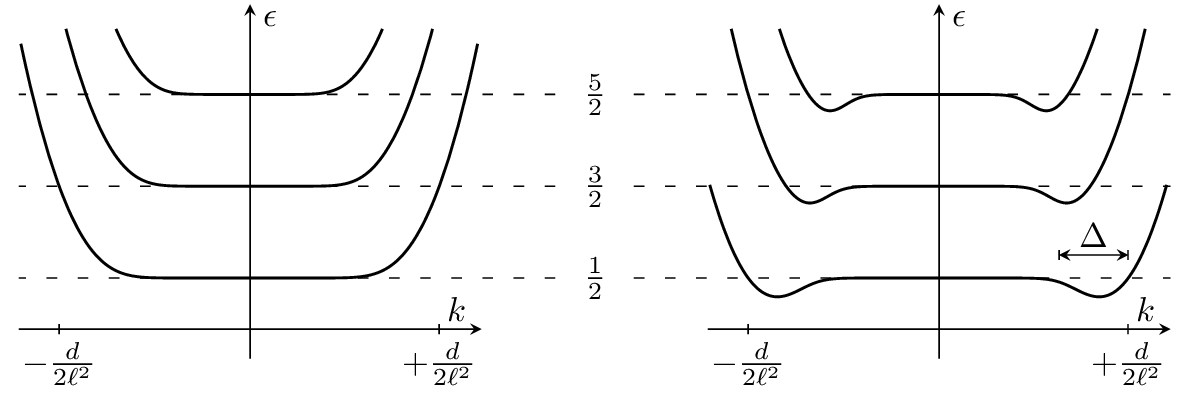}
\end{center}
\noindent
{\small Figure 1. Dispersion $\epsilon(k)$ for Dirichlet (left) and Neumann (right) BC. Shapes of dips are stable against increasing $d$.
The width of the lowest dips are of order of $\Delta\sim3\ell^{-1}$.}

\vspace*{3mm}

Dirichlet and Neumann BC produce similar flat segments in the energy curves. This feature reflects the flat structure
of the standard Landau levels where $\epsilon'(k)=0$. Distinction between the two BC arises around the segments with
nontrivial dispersion: Neumann BC cause the occurrence of dips which are absent for Dirichlet BC.
This observation is the main object of our interest.

Some remarks are in order before discussing the issue of aforementioned dips. Increasing the width $d$,
the flat segments also become wider, while the dips acquire certain stable shape. For the sake of clarity
we comment on the case of Neumann BC and consider the right dip ($k>0$).

Introduce the quantity $\kappa\equiv k\ell-\2d/\ell$ which measures the deviation of $k$ from the value of $\2d/\ell^2$.
Then the condition (15) appears as
\be
f_N(\kappa)=f_N(\kappa+d/\ell).
\ee
Provided we discuss the vicinity of $k=\2d/\ell^2$ with $d/\ell$ being large, the value of $\kappa$ is finite.
Then the right hand side of (17) can be replaced by the corresponding limit, and we come to
\be
\frac{\kappa M\big(\textstyle{\frac12}+\textstyle{\frac12}\nu,\textstyle{\frac12},-\kappa^2\big)
+2\nu\kappa M\big(\textstyle{\frac12}+\textstyle{\frac12}\nu,\textstyle{\frac32},-\kappa^2\big)}
{(1-\kappa^2)M\big(1+\textstyle{\frac12}\nu,\textstyle{\frac32},-\kappa^2\big)
+\textstyle{\frac23}(1-\nu)\kappa^2M\big(1+\textstyle{\frac12}\nu,\textstyle{\frac52},-\kappa^2\big)}
=-\frac{\Gamma(\2-\2\nu)}{\2\Gamma(-\2\nu)}.
\ee
This relation generates infinite solutions for $\nu(\kappa)$ corresponding to the dips at $k>0$.
Hereafter we discuss only the right dips since the identical analysis is valid for the left ones, as well.

Remark, that the wave functions with momenta from plateaux take nonvanishing values at
$x_L\leqslant x\leqslant x_R$, {\it i.e.} are bulk states, while those from the dips ($k\sim\pm\2\ell^{-2}d$),
are localized at boundaries thus representing the edge states. In particular, the states with $k\sim+\2\ell^{-2}d$
are localized at the left edge $x\sim-\frac12d$, and those with $k\sim-\2\ell^{-2}d$ at the right edge $x\sim+\frac12d$
(certain explicit expressions are collected in Appendix).

\section*{3. Matter Current}

Translational invariance forces the wave functions to take the form (6), and the eigenvalue problem
becomes one-dimensional on the segment $\xi_L\leqslant\xi\leqslant\xi_R$. Correspondingly, the scalar
product of two wave functions is defined as
\be
\langle\phi|\varphi\rangle=\int_{\xi_L}^{\xi_R}\phi^*(\xi)\varphi(\xi)d\xi.
\ee

Elementary calculations indicate that within the class of wave functions set either by Dirichlet or Neumann BC
we have $\langle\phi|\mathscr H\varphi\rangle=\langle\mathscr H\phi|\varphi\rangle$ signifying that
$\mathscr H$ is hermitian. Provided $\phi_k(\xi)$ is the normalized eigenfunction we have
$\epsilon(k)=\langle\phi_k|\mathscr H\phi_k\rangle$ where from we obtain
\be
\frac{d\epsilon}{dk}
=\int_{\xi_L}^{\xi_R}\phi^*\hspace*{-1.5mm}_k\frac{d\mathscr H}{dk}\phi_kd\xi
+\int_{\xi_L}^{\xi_R}\frac{d\phi^*\hspace*{-1.5mm}_k}{dk}\mathscr H\phi_kd\xi
+\int_{\xi_L}^{\xi_R}\phi^*\hspace*{-1.5mm}_k\mathscr H\frac{d\phi_k}{dk}d\xi.
\ee

Due to $\langle\phi|\mathscr H\varphi\rangle=\langle\mathscr H\phi|\varphi\rangle$ the last two terms cancel out each other and we find
\be
\frac{d\epsilon}{dk}=\ell\int_{\xi_L}^{\xi_R}\xi\hh\phi^*\hspace*{-1.5mm}_k(\xi)\phi_k(\xi)d\xi.
\ee
Comparing this to (2) we find
\be
J_y(k)=\int_{x_L}^{x_R}\mathscr J_y\hh dx=\frac{\hbar}{m\ell}\h\frac{d\epsilon}{dk},
\ee
where the left hand side is the current in $y$-direction carried by the quantum state with momentum $k$.

As a matter of (22), every quantum states with $k$ from the flat segments carries no current due to $\epsilon'(k)=0$.
The current carrying states are those with $k\sim\pm\2\ell^{-2}d$ where $\epsilon'(k)\ne0$, but the essential
difference occurs between the cases of Dirichlet and Neumann BC. Namely, in the case of Dirichlet BC, the states
with $k\sim+\2\ell^{-2}d$ carry positive current due to $\epsilon'(k)>0$, while those with $k\sim-\2\ell^{-2}d$ carry
negative current due to $\epsilon'(k)<0$.

Let us now discuss the Neumann BC and consider the states with $k\sim+\2\ell^{-2}d$ (right dip).
As already pointed out these states are all localized at the left edge. Part of them carry positive matter
current due to $\epsilon'(k)>0$, and the rest ones carry negative current due to $\epsilon'(k)<0$.
Assume now the system is filled by electrons up to the Fermi level $\epsilon_f$ as shown in figure 2 where $k_1$ and
$k_2$ are set by $\epsilon(k_1)=\epsilon(k_2)=\epsilon_f$. This is a many-body state built up of one-particle states with
$k_1\leqslant k\leqslant k_2$, hence the total current flowing along the left edge is formed by summing
up the contributions from every one-particle state involved. Combining this statement with (22) we find the
total current along the left edge vanishes (the same is true for the right edge)
\be
J_y\hspace*{-1mm}^{tot}=\int_{k_1}^{k_2}J_y(k)dk=\frac{\hbar}{m\ell}\int_{k_1}^{k_2}\frac{d\epsilon}{dk}dk
=\frac{\hbar}{m\ell}[\epsilon(k_1)-\epsilon(k_2)]=0.
\ee
Remark, that the cancellation of the total current is the precise effect.
\begin{center}
\includegraphics{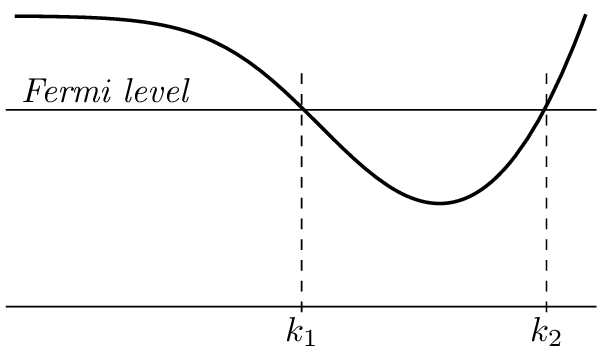}\\
\noindent{\small Figure 2. The dip filled by electrons up to the Fermi level.}
\end{center}

Summarizing, each boundary accommodates two opposite flows of matter which cancel out each other thus producing
vanishing edge currents. The precise vanishing of electric current is a good start point for developing the spin effects.
In particular, assume the electrons comprised in the many-body state shown in figure 2 are supplied with spin degree
of freedom subject to spin-orbit interaction, which roughly speaking supports the opposite spin orientations to travel
in opposite directions. In that case the electrons traveling in positive direction and those traveling in opposite direction
will carry the opposite spin orientations, {\it i.e.} the spin transport will take place with vanishing charge transport.
This is usually referred to as {\it pure spin current}. Such scenario of modelling pure spin currents is studied
in the next section.

\section*{4. Spinning electron in homogeneous orthogonal magnetic field}

We consider electrons with spin degrees of freedom exposed to Rashba interaction
(effect of Zeeman coupling is not decisive and is briefly discussed in the end).
The corresponding Hamiltonian is given by
\be
H=\frac{1}{2m}(i\hbar\partial_n+eA_n)^2-\frac{k_R}{m}\epsilon_{nj}\sigma_n(i\hbar\partial_j+eA_j),
\ee
where $k_R$ determines the scale of Rashba interaction, and $\sigma_{n=x,y}$ are Pauli matrices

Spin-orbit interaction can be incorporated into covariant derivative and (24) can be rewritten as
\babc
\begin{align}
&H=-\frac{\hbar^2}{2m}D_nD_n,\\
\nn\\
&i\hbar D_n=i\hbar\partial_n+eA_n-k_R\epsilon_{nj}\sigma_j,
\end{align}
\eabc
where the irrelevant additive constant is omitted in (25a).

\vspace*{5mm}
\noindent
\underline{\it 4.1. Charge and Spin Currents}
\vspace*{3mm}

Charge current is introduced in the standard way via the continuity equation.
Employing the Schr\"odinger equation $i\hbar\partial_t\psi=H\psi$ we obtain
$\partial_t(\psi^\dag\psi)=-\partial_n\mathscr J_n$
where from we identify the matter current to be ($n=x,y$)
\be
\mathscr J_n=\frac{\hbar}{2im}\big[\psi^\dag(D_n\psi)-(D_n\psi)^\dag\psi\big].
\ee

Analogously, we introduce the spin densities as $\psi^\dag\sigma_{\mu=x,y,z}\psi$ and arrive at
\be
\partial_t(\psi^\dag\sigma_\mu\psi)=-\h\partial_n\mathscr J_{\mu n}+T_\mu,
\ee
where the spin currents are given by
\be
\mathscr J_{\mu n}=\frac{\hbar}{2im}\big[\psi^\dag\sigma_\mu(D_n\psi)-(D_n\psi)^\dag\sigma_\mu\psi\big],
\ee
and the torques look as ($a=x,y$)
\babc
\begin{align}
T_a
&=\frac{ik_R}{m}\big[(D_a\psi)^\dag\sigma_z\psi-\psi^\dag\sigma_z(D_a\psi)\big],\\
\nn\\
T_z&=\frac{ik_R}{m}\big[\psi^\dag\sigma_n(D_n\psi)-(D_n\psi)^\dag\sigma_n\psi\big],
\end{align}
\eabc
and the summation over $n=x,y$ is implied in the last expression.

\vspace*{5mm}
\noindent
\underline{\it 4.2. Eigenvalue Problem}
\vspace*{3mm}

We use Landau gauge $\boldsymbol{A}=(0,Bx)$ with $eB<0$. Introducing
\babc
\be
\frac{1}{\ell^2}=-\frac{eB}{\hbar},
\ee
\be
\alpha=\frac{k_R\ell}{\hbar},
\ee
\be
\xi=\ell^{-1}x+k\ell,
\ee
\be
\psi(x,y)=e^{+iky}\phi(\xi),
\ee
\eabc
the eigenvalue problem for $H$ turns into $\mathscr H\phi=\epsilon\phi$ where
\be
\mathscr H
=\left\lgroup\ba{c|c}
-\2\partial_\xi^2+\2\xi^2+\alpha^2 \h\h&\h\h -\alpha(\xi+\partial_\xi)
\\\vspace*{-3.5mm}\\\hline\vspace*{-4.5mm}\\
-\alpha(\xi-\partial_\xi) \h\h&\h\h -\2\partial_\xi^2+\2\xi^2+\alpha^2
\ea\right\rgroup.
\ee

Each eigenvalue $\epsilon$ is fourfold degenerated, and the corresponding eigenstates are given by
\babc
\begin{align}
\phi_1^{}(\xi)&=e^{-\frac12\xi^2}\left\lgroup\ba{l}+2\alpha\nu_-\xi M(1-\2\nu_-,\32,\xi^2)\\\vspace*{-2mm}\\ (u-\alpha^2)M(-\2\nu_-,\2,\xi^2)\ea\right\rgroup,\\
\nn\\
\phi_2^{}(\xi)&=e^{-\frac12\xi^2}\left\lgroup\ba{l}-\alpha M(\2-\2\nu_-,\2,\xi^2)\\\vspace*{-2mm}\\ (u-\alpha^2)\xi M(\2-\2\nu_-,\32,\xi^2)\ea\right\rgroup,\\
\nn\\
\phi_3^{}(\xi)&=e^{-\frac12\xi^2}\left\lgroup\ba{l}(u+\alpha^2)\xi M(1-\2\nu_+,\32,\xi^2)\\\vspace*{-2mm}\\-\alpha M(-\2\nu_+,\2,\xi^2)\ea\right\rgroup,\\
\nn\\
\phi_4^{}(\xi)&=e^{-\frac12\xi^2}\left\lgroup\ba{l}(u+\alpha^2)M(\2-\2\nu_+,\2,\xi^2)\\\vspace*{-2mm}\\+2\alpha\nu_+\xi M(\2-\2\nu_+,\32,\xi^2)\ea\right\rgroup,
\end{align}
\eabc
where $M(a,b,z)$ is the Kummer function and
\babc
\begin{align}
\nu_\pm&=\epsilon\pm\2\sqrt{1+8\epsilon\alpha^2-4\alpha^4}\h,\\
\nn\\
u&=\2+\2\sqrt{1+8\epsilon\alpha^2-4\alpha^4}\h.
\end{align}
\eabc

General solution takes the form
\be
\phi(\xi)=c_1\phi_1(\xi)+c_2\phi_2(\xi)+c_3\phi_3(\xi)+c_4\phi_4(\xi)
\ee
and boundary conditions are necessary to fix the coefficients $c_{1,2,3,4}$.

\vspace*{5mm}
\noindent
\underline{\it 4.3. Boundary Conditions}
\vspace*{3mm}

Remind that the system under consideration is infinite in $y$ direction, but finite in $x$ direction.
Therefore one must impose some boundary conditions (BC), the physical essence of which is to prevent
the leakage of matter and spin across the edges. Assuming the system is located between $x_L=-\2d$
and $x_R=+\2d$, the boundary conditions must guarantee the $x$-components
of matter and spin currents vanish at boundaries
\be
\mathscr J_x(x=\pm\2d,y)=\mathscr J_{\mu x}(x=\pm\2d,y)=0.
\ee

We now discuss the boundary conditions. For this purpose it is reasonable to express the matter and spin currents in terms of $\phi(\xi)$.
Substituting (30e) into (26) and (28) we obtain
\babc
\begin{align}
\mathscr J_x&=\frac{\hbar}{2im\ell}\big[\phi^\dag(\nabla\phi)-(\nabla\phi)^\dag\phi\big],\\
\nn\\
\mathscr J_{\mu x}&=\frac{\hbar}{2im\ell}\big[\phi^\dag\sigma_\mu(\nabla\phi)-(\nabla\phi)^\dag\sigma_\mu\phi\big],
\end{align}
\eabc
where
\be
\nabla\phi\equiv\partial_\xi\phi+i\alpha\sigma_y\phi
=\left\lgroup\ba{l}\partial_\xi\phi_\ua+\alpha\phi_\da\\\vspace*{-2mm}\\\partial_\xi\phi_\da-\alpha\phi_\ua\ea\right\rgroup.
\ee

The currents (36) must vanish at $x=\pm\2d$. We consider two options for satisfying this requirement ($\xi_L$ and $\xi_R$ have been introduced by (9))
\begin{itemize}
\item Dirichlet BC
\be
\phi(\xi_L)=\phi(\xi_R)=0.
\ee
\item Robin BC (simultaneously involving functions and derivatives)
\be
\nabla\phi(\xi_L)=\nabla\phi(\xi_R)=0,
\ee
which for $\alpha=0$ turn into Neumann BC (involving only derivatives).
\end{itemize}

Dirichlet BC form the homogeneous system of four linear equations on the coefficients $c_{1,2,3,4}$ appearing in (34).
Solubility of the system requires the corresponding determinant vanishes. This leads to the dispersion shown
in the left panel of figure 3.

Robin BC lead to the homogeneous system of linear equations on $c_{1,2,3,4}$
and result into $\epsilon(k)$ shown in the right panel of figure 3, with still emerging dips.

\begin{center}
\includegraphics{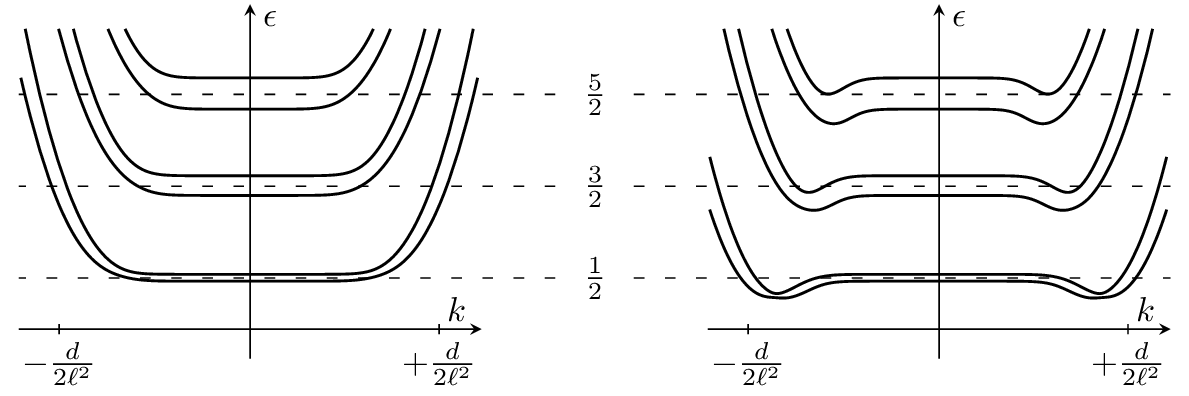}\\
\noindent
{\small Figure 3. Dispersion $\epsilon(k)$ for Dirichlet (left) and Robin (right) boundary conditions with $\alpha=0.2$.}
\end{center}

\vspace*{5mm}
\noindent
\underline{\it 4.4. Spin Currents}
\vspace*{3mm}

Assume the system is populated by electrons so that only the lowest dips are filled out, as shown in figure 4.
\begin{center}
\includegraphics{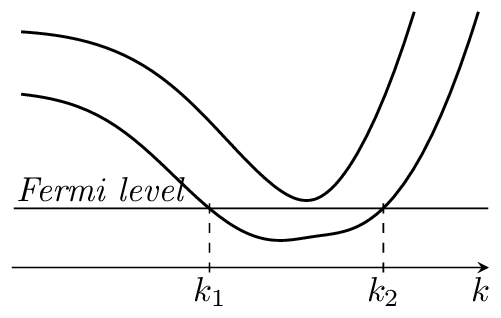}
\end{center}
\noindent
{\small Figure 4. Right dip for $\alpha=0.2$ filled up by electrons so that the Fermi level is below the next upper dip.
One-particle states with $k_1\leqslant k\leqslant k_2$ are all occupied where $k_1$ and $k_2$
are determined by $\epsilon(k_1)=\epsilon(k_2)=\epsilon_F$. Left dip is filled out in the same way.}

\vspace*{3mm}

The relation (22) derived for spinless particles holds for spinning ones as well. Therefore, in the many-body
state shown in figure 4 we again observe the precise cancellation of the charge flow independently for the left
and right edges.

We next consider the spin flow in $y$-direction.
In accord with (28) the densities of spin flow in $y$-direction appear as
\babc
\begin{align}
\mathscr J_{xy}&=\frac{\hbar}{m\ell}(\xi\phi^\dag\sigma_x\phi-\alpha\phi^\dag\phi)
=\frac{\hbar}{m\ell}\big[2\xi\phi_\ua^{}\phi_\da^{}-\alpha(\phi_\ua^2+\phi_\da^2)\big],\\
\nn\\
\mathscr J_{yy}&=\frac{\hbar}{m\ell}\h\xi\hh\phi^\dag\sigma_y\phi=0,\\
\nn\\
\mathscr J_{zy}&=\frac{\hbar}{m\ell}\h\xi\hh\phi^\dag\sigma_z\phi=\frac{\hbar}{m\ell}\h\xi(\phi_\ua^2-\phi_\da^2),
\end{align}
\eabc
where $\mathscr J_{yy}$ vanishes identically since $\phi$ is real.

Expressing the spin-orbit part of (24) as $H_{so}\sim\boldsymbol{\sigma}\cdot(\boldsymbol{p}\times\boldsymbol{E})$
we find it corresponding to the electric field $\boldsymbol{E}=(0,0,E)$. Hence, provided the electrons travel in
$y$-direction, the spin-orbit interaction is minimal when the spin points in $x$-direction. Therefore the relevant component
of the spin current must be $\mathscr J_{xy}$, representing the $y$-flow of $x$-spin.

Integrating $\mathscr J_{xy}$ over $-\2d\leqslant x\leqslant+\2d$ we obtain the $y$-flow of the $x$-spin
due to the quantum state with momentum $k$
\be
J_{xy}(k)=\int_{x_L}^{x_R}\mathscr J_{xy}(k,\xi)dx=\ell\int_{\xi_L}^{\xi_R}\mathscr J_{xy}(k,\xi)d\xi,
\ee
and the total spin current in the many-body state takes contribution from the involved one-particle states
\be
J_{xy}^{tot}=\int_{k_1}^{k_2}J_{xy}(k)dk.
\ee

Though the analytic expressions for the wave function fixed by Robin BC are available (see Appendix), the integral in (41)
can be calculated only numerically. The dependence $J_{xy}(k)$ obtained numerically for the interval of
$k$ involved in the many-body state under consideration is depicted in figure 5. From the later it is obvious that the integral
(42) takes finite value, hence the many-body state accommodates spin current accompanied by vanishing electric current.

\begin{center}
\includegraphics{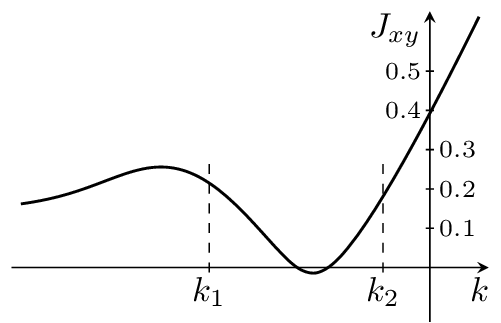}
\end{center}
\noindent
{\small Figure 5. Dimensionless current of spin $x$-component $J_{xy}$ versus $k$ for $\alpha=0.2$. The values $k_{1,2}$
are those set by $\epsilon(k_1)=\epsilon(k_2)=\epsilon_F$ in figure 4.
The corresponding value of the current of spin $z$-component $J_{zy}$ set by (40c) is of order of $10^{-14}$ which
(compared to the one of $J_{xy}$) justifies that the only non-vanishing flow of spin is the flow of its $x$-component.}

\vspace*{3mm}

Concerning the flow of $y$-spin and $z$-spin components, as already pointed out the relation $J_{yy}=0$ trivially follows from
the fact that $\phi$ is real, while the numeric calculations for $J_{zy}$ indicate it is of order of $10^{-14}$. Comparison of this
to the values of $J_{xy}$ shown in Fig. 5 justifies that the only non-vanishing flow of spin is the flow of its $x$-component.

Note that the involved one-particle states ($0<k_1\leqslant k\leqslant k_2$) are all localized at the
left edge, implying the pure spin current flows exclusively along the edge. The same is true for the right edge
($k_2\leqslant k\leqslant k_1<0$).

Summarizing, we have demonstrated that 2D system of spinning electrons subject to specific (Robin) boundary
conditions in coexistence with magnetic field and spin-orbit interaction can accommodate many-body states carrying
pure spin currents (the basic problems of generation and detection
of spin currents are beyond the scope of the given study).

Also, in our consideration the Zeeman interaction has been dropped. Would this term be included, the spins
would become canted towards $z$-direction, {\it i.e.} the non-vanishing value of $J_{zy}^{tot}$ would occur
alongside with $J_{xy}^{tot}$. However such a modification would not present any novelty and we have not discussed
this point in details.

\section*{Acknowledgments}
Authors would like to thank A. A. Nersesyan, G. I. Japaridze and M. Sekania for fruitful discussions and remarks.
G. Ts. is grateful to C. Morais Smith, C. Ortix and G. van Miert for helpful discussions during
the stay at Utrecht University. The work was supported by Rustaveli National Science Foundation through grant
FR/265/6-100/14.

\section*{Appendix}

\def\theequation{A.\arabic{equation}}
\setcounter{equation}{0}

Kummer functions appearing in eigenstates (32) comprise exponential divergencies for large arguments, and therefore
are not appropriate for numeric calculation. In order to avoid this drawback we introduce the combinations
\babc
\begin{align}
P(\nu,\xi)&\equiv e^{-\xi^2}\left[\frac{M(-\2\nu,\2,\xi^2)}{\Gamma(\2-\2\nu)}+\frac{2|\xi|M(\2-\2\nu,\32,\xi^2)}{\Gamma(-\2\nu)}\right]
=\frac{M(\2+\2\nu,\2,-\xi^2)}{\Gamma(\2-\2\nu)}+\frac{2|\xi|M(1+\2\nu,\32,-\xi^2)}{\Gamma(-\2\nu)},\\
\nn\\
Q(\nu,\xi)&\equiv\frac{M(-\2\nu,\2,\xi^2)}{\Gamma(\2-\2\nu)}-\frac{2|\xi|M(\2-\2\nu,\32,\xi^2)}{\Gamma(-\2\nu)}=\frac{U(-\2\nu,\2,\xi^2)}{\sqrt{\pi}},
\end{align}
\eabc
where $U(a,b,z)$ is Tricomi function.

Advantage of employing these combinations is that all exponential factors contained
in $M$ are explicitly taken into account, and $P(\nu,\xi)$ and $Q(\nu,\xi)$ behave at $|\xi|\to\infty$ as follows
\babc
\begin{align}
P(\nu,\xi)&\to\frac{-2\hh{\tt sin}(\pi\nu)}{\pi}\sum_{s=0}^\infty\frac{\Gamma(2s+1+\nu)}{|2\xi|^{2s+1+\nu}s!},\\
\nn\\
Q(\nu,\xi)&\to\frac{|\xi|^\nu}{\sqrt{\pi}\h\Gamma(-\nu)}\sum_{s=0}^\infty\frac{(-1)^s\Gamma(2s-\nu)}{|2\xi|^{2s}s!}.
\end{align}
\eabc

Employing $P(\nu,\xi)$ and $Q(\nu,\xi)$ below we present the explicit expressions for the wave function (34) where the
coefficients $c_{1,2,3,4}$ are fixed by the Robin boundary conditions (39). We are interested in the values
of $k$ from the right dip $k\sim+\2\ell^{-2}d$, provided the width $d$ is large and separately consider
$k\lesssim+\2\ell^{-2}d$ and $k\gtrsim+\2\ell^{-2}d$.

For $k\lesssim+\2\ell^{-2}d$ the wave function (not normalized) takes the form
\babc
\begin{align}
\phi(0\leqslant\xi\leqslant\xi_R)&=e^{-\frac12\xi^2}\left\lgroup\ba{l}
(u+\alpha^2)\hh AQ(\nu_+-1,\xi)-\alpha\nu_-BQ(\nu_--1,\xi)
\\\\
2\hh \alpha\hh AQ(\nu_+,\xi)+(u-\alpha^2)BQ(\nu_-,\xi)
\ea\right\rgroup,\\
\phi(\xi_L\leqslant\xi\leqslant 0)&=e^{+\frac12\xi^2}\left\lgroup\ba{l}
(u+\alpha^2)\hh AP(\nu_+-1,\xi)-\alpha\nu_-BP(\nu_--1,\xi)
\\\\
2\hh \alpha\hh AP(\nu_+,\xi)+(u-\alpha^2)BP(\nu_-,\xi)
\ea\right\rgroup,
\end{align}
\eabc
where
\begin{align}
A&=u\big[-(1-u^{-1}\alpha^2)|\xi_L|P(\nu_-,\xi_L)-\nu_-P(\nu_--1,\xi_L)\big],\\
\nn\\
B&=\alpha\big[2|\xi_L|P(\nu_+,\xi_L)-(u+\alpha^2-2\nu_+)P(\nu_+-1,\xi_L)\big].
\end{align}
Remark, that $\xi_L\equiv-\2\ell^{-1}d+kl$ is small and negative here, while $\xi_R\equiv+\2\ell^{-1}d+kl$ is positive and large,
hence $\phi(\xi)$ set by (A.3) is localized at the left edge ($\xi\sim\xi_L$).

For $k\gtrsim+\2\ell^{-2}d$ we have $0<\xi_L\leqslant\xi\leqslant\xi_R$ and the wave function is expressed by unique expression
\be
\phi(\xi)
=e^{-\frac12\xi^2}\left\lgroup\ba{l}
(u+\alpha^2)\hh CQ(\nu_+-1,\xi)-\alpha\nu_-DQ(\nu_--1,\xi)
\\\\
2\hh \alpha\hh CQ(\nu_+,\xi)+(u-\alpha^2)\hh DQ(\nu_-,\xi)
\ea\right\rgroup,
\ee
where
\babc
\begin{align}
C&=\alpha\big\{(u-\alpha^2+2\nu_-)Q(\nu_-,\xi_L)-\nu_-|\xi_L|Q(\nu_--1,\xi_L)\big\},\\
\nn\\
D&=u\big\{2Q(\nu_+,\xi_L)-(1+u^{-1}\alpha^2)|\xi_L|Q(\nu_+-1,\xi_L)\big\}.
\end{align}
\eabc
Provided $\xi_L$ is small and $\xi_R$ is large, the wave function is localized at the left edge.

\end{document}